\documentclass[prd,preprint,superscriptaddress,showpacs,nofootinbib,%
tightenlines ]{revtex4}
\usepackage{graphics}
\usepackage{epsfig}
\usepackage{amssymb,amsmath}
\newcommand{\ben}{\begin{displaymath}}
\newcommand{\een}{\end{displaymath}}
\newcommand{\be}{\begin{equation}}
\newcommand{\ee}{\end{equation}}
\newcommand{\bea}{\begin{eqnarray}}
\newcommand{\eea}{\end{eqnarray}}

\begin{document}
\title{Complex-mass renormalization in hadronic EFT:\\ applicability at two-loop order\\[0.5em]}
\author{D.~Djukanovic}
\affiliation{Helmholtz Institute Mainz, University of Mainz, D-55099 Mainz, Germany}
\author{E.~Epelbaum}
 \affiliation{Institut f\"ur Theoretische Physik II, Ruhr-Universit\"at Bochum,
D-44780 Bochum, Germany}
\author{J.~Gegelia}
\affiliation{Institute for Advanced Simulation, Institut f\"ur Kernphysik
   and J\"ulich Center for Hadron Physics, Forschungszentrum J\"ulich, D-52425 J\"ulich,
Germany}
  \affiliation{Tbilisi State  University,  0186 Tbilisi,
 Georgia}
\author{H.~Krebs}
 \affiliation{Institut f\"ur Theoretische Physik II, Ruhr-Universit\"at Bochum,  D-44780 Bochum,
 Germany}
\author{U.-G.~Mei\ss ner}
 \affiliation{Helmholtz Institut f\"ur Strahlen- und Kernphysik and Bethe
   Center for Theoretical Physics, Universit\"at Bonn, D-53115 Bonn, Germany}
 \affiliation{Institute for Advanced Simulation, Institut f\"ur Kernphysik
   and J\"ulich Center for Hadron Physics, Forschungszentrum J\"ulich, D-52425 J\"ulich,
Germany}
\begin{abstract}
   We discuss the application of the complex-mass scheme
to multi-loop diagrams in hadronic effective field theory by considering as
an example a two-loop self-energy
diagram.  We show that the renormalized two-loop diagram satisfies the
power counting.
\end{abstract}
\date{June 12, 2015}
\maketitle
\section{Introduction\label{introduction}}

The extension of  mesonic chiral perturbation theory
\cite{Weinberg:1979kz,Gasser:1984gg,Gasser:1984yg} to include
heavier (non-Goldstone) degrees of freedom is known to be a non-trivial problem.
Already in the one-nucleon sector it was found that higher-order
loops contribute to lower-order calculations \cite{Gasser:1988rb}.
This problem, also for the delta resonance included as a dynamical degree of freedom,
has been solved in the framework of the
heavy-baryon chiral perturbation theory
\cite{Jenkins:1991jv,Bernard:1992qa,Hemmert:1996xg} and
later in the original manifestly Lorentz-invariant formulations by using
the infrared regularization and
the extended on-mass-shell renormalization (EOMS)
\cite{Tang:1996ca,Ellis:1997kc,Becher:1999he,Gegelia:1999gf,Gegelia:1999qt,Fuchs:2003qc}.
It is also possible to consistently
include virtual (axial-) vector mesons
in effective field theory (EFT) \cite{Kubis:2000zd,Fuchs:2003sh,Schindler:2003xv}
for processes
involving soft external pions and nucleons with small three-momenta.
On the other hand, as the (axial-) vector mesons decay into light
modes (and therefore large imaginary parts appear),
the issue of including (axial-) vector mesons
in an EFT for energies when the intermediate resonant
states can be generated is still very problematic \cite{Bruns:2004tj}.
First attempts have been made to handle this problem by applying the
complex mass scheme \cite{Stuart:1990,Denner:1999gp} in
Refs.~\cite{Djukanovic:2009zn,Djukanovic:2009gt,Djukanovic:2010id,Bauer:2011bv,Bauer:2012at,Djukanovic:2013mka}.\footnote{Note that the very non-trivial issue of unitarity within
the complex-mass scheme has been addressed
in Ref.~\cite{Bauer:2012gn} and recently it  has
been thoroughly investigated in Ref.~\cite{Denner:2014zga}.}

The aim of this work is to explicitly demonstrate the applicability of
the complex-mass scheme to multi-loop diagrams.
For that we analyze  a two-loop self-energy diagram within the
complex-mass scheme, similar to what was done in
Ref.~\cite{Schindler:2003je} using the
infrared regularization and the EOMS renormalization.
We show that the resulting renormalized expressions for the two-loop
diagram indeed satisfy the  power counting of the considered EFT.

\section{A two-loop $\omega$-meson self-energy diagram}

\begin{figure}
\epsfig{file=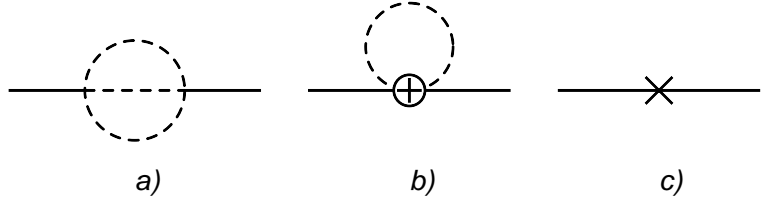,width=10cm} \caption[]{\label{fse:fig}
Two-loop order self-energy diagrams. The vertex with a circled cross  corresponds to a one-loop order counterterm
and a mere cross corresponds to two-loop order
counterterms. Solid/dashed lines correspond to heavy/light particles.}
\end{figure}

We start with a two-loop self-energy diagram of the $\omega$-meson shown in
Fig.~\ref{fse:fig}~a), where the solid lines correspond to the vector meson and the
dashed ones to the pion. This diagram is generated by the interaction
Lagrangian \cite{Meissner:1987ge}:
\begin{equation}
{\cal L}_{V\Phi\Phi\Phi}^{(1)}=\frac{i\,h}{4 F_\pi^3}\,\epsilon^{\mu\nu\alpha\beta}
\,{\rm tr}\left( V_\mu \partial_\nu\Phi \partial_\alpha\Phi\partial_\beta\Phi\right),
\end{equation}
where $V_\mu$ is the vector field corresponding to the $\omega$ meson and
$\Phi= \pi^a \tau^a$ to the pions (for the case of
the two-flavor chiral effective field theory, which we consider here).

Calculating the diagram of Fig.~\ref{fse:fig}~a) we obtain:
\begin{equation}
i\,\Sigma^{\mu\nu} = \frac{9\,i h^2}{F_\pi^6} \,\epsilon^{\mu\lambda_1\sigma_1\kappa_1}\,
\epsilon^{\nu\lambda_2\sigma_2\kappa_2}\,p_{\kappa_1} p_{\kappa_2} I_{\lambda_1\sigma_1\lambda_2\sigma_2},
\label{oSE}
\end{equation}
where
\begin{equation}
I^{\lambda_1\sigma_1\lambda_2\sigma_2}=\frac{1}{(2\pi)^{2 n}}\int\frac{d^nk_1 d^nk_2\,k_1^{\lambda_1}k_1^{\lambda_2}k_2^{\sigma_1}k_2^{\sigma_2}}{\left[k_1^2-M^2+i 0^+\right]
\left[k_2^2-M^2+i 0^+\right]\left[(p+k_1+k_2)^2-M^2+i 0^+\right]}.
\label{deftint}
\end{equation}
Using the tensor integral reduction formulas specified in the Appendix~C
of Ref.~\cite{Schindler:2007dr} and simplifying Eq.~(\ref{oSE}) we obtain
\begin{equation}
i\,\Sigma^{\mu\nu} = \frac{72\,i \pi^2 h^2}{F_\pi^6}\,\left(p^\mu p^\nu-p^2 g^{\mu\nu}\right) I_{n+2},
\label{oSER}
\end{equation}
where
\begin{equation}
I_{n+2}=\frac{1}{(2\pi)^{2 (n+2)}}\int\frac{d^{n+2}k_1 d^{n+2}k_2}{\left[k_1^2-M^2+i 0^+\right]
\left[k_2^2-M^2+i 0^+\right]\left[(p+k_1+k_2)^2-M^2+i 0^+\right]}.
\label{deftintsca} 
\end{equation}
Thus the investigation of the considered two-loop diagram  contributing
to the $\omega$-meson self-energy
reduces to the study of a scalar integral in 6 dimensions.

\section{Lagrangian and power counting\label{lpc}}

To properly subtract the two-loop integral of the previous section by applying
the complex mass scheme  without unnecessary complications due to the spin and
chiral structure of the low-energy effective field theory,
we focus here on an effective field theoretical model Lagrangian of interacting
scalar fields in six space-time dimensions
\begin{equation}\label{lagrange}
\mathcal{L}=\frac{1}{2}(\partial_{\mu}\pi\partial^{\mu}\pi-M^2\pi^2)+
\frac{1}{2}(\partial_{\mu}\Psi\partial^{\mu}\Psi-m^2\Psi^2)-
\frac{g}{3!}\,\pi^3\Psi +\mathcal{L}_1~,
\end{equation}
where the masses of the scalar fields $\pi$ and $\Psi$ satisfy the condition
$M\ll m$ (i.e. $\Psi$ represents an unstable particle).
   The Lagrangian $\mathcal{L}_1$ contains all possible terms which are
consistent with the Lorentz symmetry and with the invariance under the simultaneous
transformations $\pi\to -\pi$ and $\Psi \to -\Psi$.
Within the considered EFT, we drop
heavy-particle loops, however, we compensate for their contributions
by including them in the low-energy constants.
The application of the complex-mass scheme guarantees that
subtracted Feynman diagrams have a certain ``chiral'' order $D$,
specified by the power counting. In particular, let $Q$ stand for
small quantities like the mass $M$, small external four-momenta of $\pi$ or
small external three-momenta of $\Psi$. Then
the vertex generated by the $\pi\Psi$ interaction explicitly shown in
Eq.~(\ref{lagrange}) counts as $Q^0$, the $\Psi$ propagator as $Q^{-1}$,
the $\pi$ propagator as $Q^{-2}$, and a loop integration in $n$ dimensions
as $Q^n$, respectively.

\section{Application to the two-loop self-energy\label{application}}

Here we consider the $\Psi$
self-energy diagram shown in Fig.~\ref{fse:fig}~ a). The corresponding expression reads:
\begin{equation}\label{selfenergy}
-i\Sigma_{\Psi}(p)=\frac{i\, g^2}{6 (2\pi)^{2n}} \int\hspace{-2mm}\int
\frac{d^nk_1d^nk_2}
{(k_1^2-M^2+i0^+)(k_2^2-M^2+i0^+)[(p+k_1+k_2)^2-M^2+i0^+]}~ ,
\end{equation}
where $1/6$ is a symmetry factor and
$n$ denotes the number of space-time dimensions.
   According to the above power counting, the
diagram of Fig.~\ref{fse:fig}~a) has the order $Q^{2n-4}$.\footnote{Note that one of the $\pi$-propagators
carries a large momentum and thus counts as $\mathcal{O}(Q^0)$.}

   Using the dimensional counting analysis of Ref.~\cite{Gegelia:zz}
(or equivalently, the ''strategy of regions'' \cite{Beneke:1997zp}),
the self-energy $\Sigma_{\Psi}$ for $M\to 0$ and $p^2\sim m^2$ can be written as
\begin{equation}\label{intsplit}
\Sigma_{\Psi}=F(p^2,M^2,n) + M^{n-2}G(p^2,M^2,n)+M^{2n-4}H(p^2,M^2,n)~,
\end{equation}
where the functions $F(p^2,M^2,n)$, $G(p^2,M^2,n)$, and
$H(p^2,M^2,n)$ can be expanded in non-negative integer powers of
$M^2$. The coefficients of the Taylor expansion of $F$ in $M^2$ can be obtained
by expanding the integrand
of Eq.~(\ref{selfenergy}) in $M^2$ and interchanging summation and integration:
\begin{equation}\label{F}
F(p^2,M^2,n)=-\frac{g^2}{6}\sum_{i,j,l=0}^{\infty}
\frac{(M^2)^{i+j+l}}{(2\pi)^{2n}}
\int\hspace{-2mm}\int\hspace{-1mm}\frac{d^nk_1d^nk_2}
{(k_1^2+i0^+)^{1+i}(k_2^2+i0^+)^{1+j}[(p+k_1+k_2)^2+i0^+]^{1+\,l}}~.
\end{equation}
Calculating the integrals of Eq.~(\ref{F}) we obtain
\begin{eqnarray}\label{Fcoeff}
&&\int\hspace{-2mm}\int\frac{d^nk_1d^nk_2}
{(k_1^2+i0^+)^{1+i}(k_2^2+i0^+)^{1+j}[(p+k_1+k_2)^2+i0^+]^{1+ \,l}}= \pi ^n i^{-2 (i+j+l)} \left(-p^2-i 0^+\right)^{n-i-j-l-3}  \nonumber\\
&& \times \frac{ \Gamma
   \left(\frac{n}{2}-i-1\right) \Gamma
   \left(\frac{n}{2}-j-1\right) \Gamma
   \left(\frac{n}{2}-l-1\right) \Gamma
   (i+j+l-n+3)}{\Gamma (i+1) \Gamma (j+1) \Gamma (l+1) \Gamma
   \left(\frac{3 n}{2}-i-j-l-3\right)}~.
\end{eqnarray}
The terms in Eq.~(\ref{F}) are analytic in $M^2$.  In order to find the power
counting violating terms, we need to expand the coefficients of
the series in Eq.~(\ref{F}) in terms of $p^2-m^2$. We only need to subtract
those terms which violate the power counting, which in the present case
of six space-time dimensions are all terms of order $Q^7$ or less. Doing so we
identify the subtraction terms which are analytic in $M^2$ and $p^2-M^2$.
These subtraction terms are canceled by counterterm contributions, shown in
Fig.~\ref{fse:fig}~c). As is clear from Eq.~(\ref{Fcoeff}), the counterterms
have to be complex.

\medskip

   Next we investigate $G(p^2,M^2,n)$ which can be found from
Eq.~(\ref{selfenergy}) as  a sum of terms identified by re-scaling
($k_1\mapsto M k_1, k_2\mapsto k_2$), ($k_1\mapsto k_1, k_2\mapsto Mk_2$) and ($k_1\mapsto M k_1-k_2-p, k_2\mapsto k_2$).
In all cases, the re-scaling generates an overall factor of $M^{n-2}$.
The remaining integrands need to be expanded in $M$, the summation
and integration interchanged. Doing the above manipulations and adding all terms together we obtain
\begin{equation}
\label{G}
G(p^2,M^2,n)=-\frac{g^2}{2\,(2\pi)^{2n}}\sum_{i,j=0}^{\infty}
\sum_{a=0}^{j}\sum_{b=0}^{a}
(-1)^j2^{j-b}
\left(\begin{array}{c}j\\a\end{array}\right)
\left(\begin{array}{c}a\\b\end{array}\right)
M^{2i+j+b}
I_{ij,ab}(p^2,n)~,
\end{equation}
with the binomial coefficients
\begin{displaymath}
\left(\begin{array}{c}r\\s\end{array}\right)
=\frac{r!}{s!(r-s)!}~,
\end{displaymath}
and the integrals $I_{ij,ab}(p^2,n)$ given as
\begin{equation}
\label{loopint}
I_{ij,ab}(p^2,n)=\int\hspace{-2mm}\int
\frac{d^nk_1d^nk_2\,(p\cdot k_1)^{j-a}(k_1\cdot k_2)^{a-b}(k_1^2-1)^b}
{(k_1^2-1+i0^+){(k_2^2+i0^+)}^{1+i}{[(p+k_2)^2+i0^+]}^{1+j}}~.
\end{equation}
The non-vanishing terms in Eq.~(\ref{G}) contain only nonnegative integer
powers of $M^2$. This is because for $j+b$ odd, the loop integral of
Eq.~(\ref{loopint}) is an odd function of $k_1$ and hence vanishes.

\begin{figure}
\epsfig{file=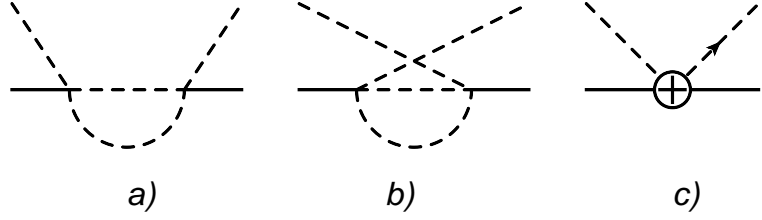,width=10cm} \caption[]{\label{scatt:fig}
One-loop diagrams contributing to $\pi\Psi$ scattering.}
\end{figure}

The only power counting violating contributions are contained in the terms of
Eq.~(\ref{G}) with $2 i+j+b$ equal to either $0$ or $2$.
Calculating these contributions we obtain:
\begin{eqnarray}
\label{loopint0}
M^{n-2} \Delta G & = & \frac{g^2 M^{n-2} \left(-m^2-i 0^+\right)^{\frac{n}{2}-5} }{96 (4 \pi)^n}
   \Gamma
   \left(1-\frac{n}{2}\right) \Gamma \left(2-\frac{n}{2}\right)
   \Gamma \left(\frac{n}{2}-1\right)^2 \Biggl\{\frac{48\,m^6}{\Gamma (n-2)}
   \nonumber\\
&+& 6 m^2
   \left[\frac{\left(m^2-p^2\right)^2
   \left(n-4\right)(n-6)}{\Gamma (n-2)}-\frac{8 m^2 M^2 (n+4)
   }{n \Gamma (n-3)}\right]- \frac{24 m^4 \left(m^2-p^2\right)
   \left(n-4\right)}{\Gamma (n-2)}\nonumber\\
&-& (n-6)
   \left(m^2-p^2\right) \left(\frac{\left(m^2-p^2\right)^2
   \left(n-4\right)(n-8)}{\Gamma (n-2)}-\frac{24 m^2
   M^2 (n+4)}{n \Gamma
   (n-3)}\right)\Biggr\}.
\end{eqnarray}
These terms are canceled by the diagram shown in Fig.~\ref{fse:fig}~b).
The one-loop counterterms contributing in this diagram originate from
the one-loop diagrams of Fig.~\ref{scatt:fig}~a) and b). Expressions corresponding
to these diagrams are given by
\begin{equation}\label{sub1}
\mathcal{M}_{2a}=
\frac{g^2}{2} \int\frac{d^nk}{(2\pi)^n}\frac{1}{(k^2-M^2+i0^+)[(k+p+q)^2-M^2+i0^+]}~,
\end{equation}
and
\begin{equation}\label{sub2}
\mathcal{M}_{2b}=
\frac{g^2}{2} \int\frac{d^nk}{(2\pi)^n}\frac{1}{(k^2-M^2+i0^+)[(k+p-q')^2-M^2+i0^+]}~,
\end{equation}
respectively.

\noindent
The diagrams in Fig.~\ref{scatt:fig}~a) and b) are of the order $Q^{n-2}$.
Loop integrals in Eqs.~(\ref{sub1}) and (\ref{sub2})
contain contributions that violate the power counting, namely:
\begin{eqnarray}
\label{deleoms}
\Delta\mathcal{M}_{2a}+\Delta\mathcal{M}_{2b}
&=& \frac{ i g^2 \lambda(m,n) }{48 m^6 (n-4)} \biggl\{2 m^6 (n-10) (n-8) (n-6)\nonumber\\
&-&3 m^4 (n-8) \left(32 M^2
   (n-3)+(n-10) (n-4) \left[(p-q')^2+(p+q)^2\right]\right)\nonumber\\
&+& 3 m^2 (n-6)
   \biggl[16 M^2 (n-3) \left[(p-q')^2+(p+q)^2\right]\nonumber\\
&+& (n-10) (n-4)
   \left(\left((p-q')^2\right)^2+\left((p+q)^2\right)^2\right)\biggr]\nonumber\\
&-& (n-8) (n-6) (n-4)
   \left[\left((p-q')^2\right)^3+\left((p+q)^2\right)^3\right]
   \biggr\}, \nonumber\\
\lambda(m,n) &=& \frac{\Gamma
   \left(3-\frac{n}{2}\right) \Gamma
   \left(\frac{n}{2}-1\right)^2
   }{ (4 \pi)^{n/2} \Gamma (n-2)} \left(-\frac{m^2+i0^+}{\mu^2}\right)^{n/2-2}~.
\end{eqnarray}
The counterterm diagram which cancels these contributions is shown in
Fig.~\ref{scatt:fig}~c) and is generated by {a rather complicated Lagrangian}
which is contained in the ${\cal L}_1$ term of Eq.~(\ref{lagrange}).

When calculating  the self-energy of $\Psi$, these counterterms give a
contribution shown in Fig.~\ref{fse:fig}~b).
The corresponding expression reads
\begin{eqnarray}\label{CTselfir}
-i\Sigma_{CT}
&=&-\frac{ig^2\lambda(m,n)}{48 m^6 (n-4) n}\Biggl\{m^6 (n-10) (n-8) (n-6) n \nonumber\\
&-&3 m^4 (n-8) n \left[M^2 (n-2)
   (n+4)+(n-10) (n-4) p^2\right] \nonumber\\
&+& 3 m^2 (n-6) \biggl[M^4 (n-2) n
   (n+4)  \nonumber\\
&+& 2 M^2 (n+4) ((n-8) n+20) p^2+(n-10) (n-4) n
   p^4\biggr]\nonumber\\
&-& (n-8) (n-6) (n-4) \left(M^2+p^2\right) \left(M^4 n+2
   M^2 (n+6) p^2+n p^4\right)
\Biggr\}I_{\pi},
\end{eqnarray}
where
\begin{equation}\label{Ipi}
I_{\pi}=i\int\frac{d^nk}{(2\pi)^n}\frac{1}{k^2-M^2+i0^+} =
M^{n-2}\frac{\Gamma(1-n/2)}{(4\pi)^{n/2}}~.
\end{equation}
Indeed, as already mentioned above, the self-energy contribution of
Eq.~(\ref{CTselfir}) cancels the power counting violating contributions in Eq.~(\ref{loopint0}).

   { Finally, the function $H(p^2,M^2,n)$ is given as a sum of three terms obtained from Eq.~(\ref{selfenergy}) by
re-scalings ($k_1\mapsto Mk_1$, $k_2\mapsto Mk_2$), ($k_1\mapsto Mk_1-p$, $k_2\mapsto Mk_2$) and ($k_1\mapsto Mk_1$, $k_2\mapsto Mk_2-p$),
extracting a factor of $M^{2n-4}$, expanding the remaining integrand
in $M$, and interchanging integration and summation, yielding
\begin{eqnarray}
\label{H}
H(p^2,M^2,n)&=&-\frac{g^2}{2(2\pi)^{2n}}
\frac{1}{p^2}\sum_{i=0}^{\infty}
(-1)^i\left(\frac{M}{p^2}\right)^i\nonumber\\
&&\times \sum_{j=0}^i
\left(
\begin{array}{c}
i\\
j
\end{array}
\right)
\int\hspace{-2mm}\int d^nk_1d^nk_2
\frac{[2p\cdot(k_1+k_2)]^{i-j}[M(k_1+k_2)^2-M]^j}{(k_1^2-1+i0^+)
(k_2^2-1+i0^+)}~.
\end{eqnarray}}
It is easy to see that
Eq.~(\ref{H}), in combination with the factor $M^{2n-4}$ of
Eq.~(\ref{intsplit}), satisfies the power counting.

   Combining the results above, we conclude that all terms violating the power counting
are canceled in the sum of diagrams shown in Fig.~\ref{fse:fig}.

\section{Conclusion\label{conclusion}}
   In conclusion, using as example a two-loop self-energy diagram in an EFT of
interacting light and heavy
scalar particles we have demonstrated that the application of the complex-mass scheme to two-loop
diagrams leads to a consistent power counting.
As expected from  general considerations, the subtraction of one-loop sub-diagrams
plays an important role in the renormalization of the two-loop diagrams.
Our example provides an explicit illustration
of the fact that the application of the complex-mass scheme to
multi-loop diagrams of a low-energy effective field theory leads to a
consistent power counting.
   Calculations using the chiral effective Lagrangians are
more involved due to the complicated structure of the interactions, but
the general features of the renormalization program
do not change. As we have shown, such a typical heavy-light two-loop self-energy
diagram emerges naturally when considering a chiral EFT with $\omega$-mesons and pions.

\acknowledgments
This work was supported in part by Georgian Shota Rustaveli National
Science Foundation (grant FR/417/6-100/14),
DFG (SFB/TR 16, ``Subnuclear Structure of Matter''),
and ERC project 259218 NUCLEAREFT.

\end{document}